\documentclass{elsart}

\usepackage{amssymb}

\begin{document}

\begin{frontmatter}

\title{The redshifts of bright sub-mm sources}

\author{James S. Dunlop}

\address{Institute for Astronomy, University of Edinburgh, Royal
Observatory, Edinburgh EH9 3HJ, UK.}

\begin{abstract}
One of the key goals in observational cosmology over the next few years will 
be to establish the redshift distribution of the recently-discovered sub-mm 
source population. In this brief review I discuss and summarize the redshift 
information which has been gleaned to date for the $\simeq 50$ bright sub-mm 
sources which have been uncovered via the six main classes of survey performed 
with SCUBA on the JCMT over the last 2-3 years. Despite the biases inherent in 
some of these surveys, and the crudeness of the redshift information available 
in others, I conclude that all current information suggests that only 
$10-15$\% of luminous sub-mm sources lie at $z < 2$, and that the median 
redshift of this population is $z \simeq 3$. I suggest that such a high median 
redshift is arguably not unexpected given current theories designed to explain 
the correlation between black-hole mass and spheroid mass found at low 
redshift. In such scenarios, peak AGN emission is expected to correspond to, 
or even to cause termination of major star-formation activity in the host 
spheroid. In contrast, maximum dust emission is expected to occur roughly 
half-way through the star-formation process. Given that optical emission from 
bright quasars peaks at $z \simeq 2.5$, dust-emission from massive ellipticals 
might be reasonably expected to peak at some point in the preceding $\simeq 1$ 
Gyr, at $z \simeq 3$. Confirmation or refutation of this picture requires 
significantly-improved redshift information on bright samples of SCUBA sources.
\end{abstract}

\end{frontmatter}

\section{Introduction}
\vspace*{-0.5cm}
The importance of understanding the nature
of the recently-discovered sub-mm galaxy population has been
discussed in detail by many authors (e.g. Hughes et al. 1998; Blain et
al. 1999). Most importantly,
constraining the redshift distribution of this population
is a requirement for obtaining
an unbiased view of the star-formation history of the universe, and 
potentially holds the key to understanding the formation and evolution of 
massive elliptical galaxies (e.g. Eales et al. 2000; Dunlop 2001). 
In this review I therefore 
concentrate on what we have learned to date about the redshift distribution
of the sub-mm population from the six main classes of survey which researchers
have undertaken in the last 2-3 years with the SCUBA camera on the JCMT.

First I consider current results from three different types of surveys aimed
at determining the sub-mm luminosities of objects which have already been 
discovered at other wavelengths. These are i) radio galaxies, 
ii) optically selected quasars, and iii) micro-Jy radio sources. Such 
studies have the practical advantage of accurate source positions, and 
(at least
in the first 2 cases) pre-existing redshift information. By focussing
on specific sub-classes of object they are inevitably biased to some extent,
but the extent of this bias may not necessarily be severe.

Second, I consider current results from surveys which have been designed
to discover new sources at sub-mm wavelengths. Again, I have divided
these into three distinct classes, namely i) surveys which aim to 
take advantage of the gravitational lensing effects of intervening rich 
clusters, ii) imaging surveys of the fields around known high-redshift
objects, and iii) genuinely unbiased, blank-field surveys of non-descript
regions of sky.

For the sake of clarity and comparability I have produced summary statistics
for each survey considering only sources which have been detected in the 
sub-mm at significance levels $> 4\sigma$, and which have intrinsic sub-mm
flux densities $S_{850} > 4$mJy.

Before proceeding with this census, it
is important to note that, apart from the radio galaxies and quasars 
for which spectroscopic redshifts already exist, almost all of the redshift
estimates discussed and summarized in what follows are based on the 
observed far-infrared$\rightarrow$radio spectral energy distributions (SEDs)
of the sources. This serves to emphasize the importance of
understanding such SEDs, and legitimizes the inclusion of this review in 
this particular conference! However it does of course mean that many of the 
current redshift constraints are inevitably rather crude and uncertain.

Of greatest current importance is the use of the sub-mm:radio 
flux-density ratio $S_{850\mu m}:S_{1.4GHz}$ which is a relatively sensitive
function of redshift for a starburst galaxy, initially rising 
$\propto (1+z)^{4.5}$ before flattening towards higher redshift in a manner
which depends on the form of the assumed template spectrum. This ratio
was first used by Hughes et al. (1998) to set redshift constraints on the 
sub-mm sources detected in the Hubble Deep Field, and has since 
been explored in detail by Carilli \& Yun (1999), Carrilli \& Yun (2000),
Dunne et al. (2000), Barger et al. (2000), and Blain (2001).
There is now general agreement that adopting M82 as a template SED 
(e.g. Carilli \& Yun 1999) is probably {\it not} appropriate for the 
luminous sub-mm population and is likely to lead to excessive 
redshift estimates. Instead, most authors now adopt Arp220 as the most appropriate 
low-redshift template ULIRG. 
Alternatively,
Dunne et al. (2000) have chosen to derive a mean-SED from their SCUBA 
survey of low-redshift IRAS galaxies. There
is some continuing controversy over 
the precise form of the best relation to use. Dunne et al. (2000) argue 
convincingly that their relation is most robustly calibrated at low redshift. 
On the other hand the revised relation presented by Carilli \& Yun (2000) 
appears to do a better job of reproducing the redshifts of the handful of 
objects
with known CO redshifts at $z > 2$. However, the level of this disagreement is
now comparable to the uncertainties in these relations, and the key point
in what follows is that essentially all authors in this field would 
appear to agree that a flux-density ratio $S_{850}:S_{1.4GHz} > 100$
(or equivalently a spectral index $\alpha^{850}_{1.4} > 0.84$) implies
$z > 2$. Finally, when utilising this redshift-estimation technique,
it is always worth remembering that over-optimistic radio-submm associations, 
or failure to recognize an AGN
contribution to radio flux-density, will always result in an 
{\it under-estimate} of the true redshift of the source. Over-estimation
of redshift is less likely to occur.

\begin{figure}
\vspace{5cm}
\centering
\setlength{\unitlength}{1mm}
\includegraphics{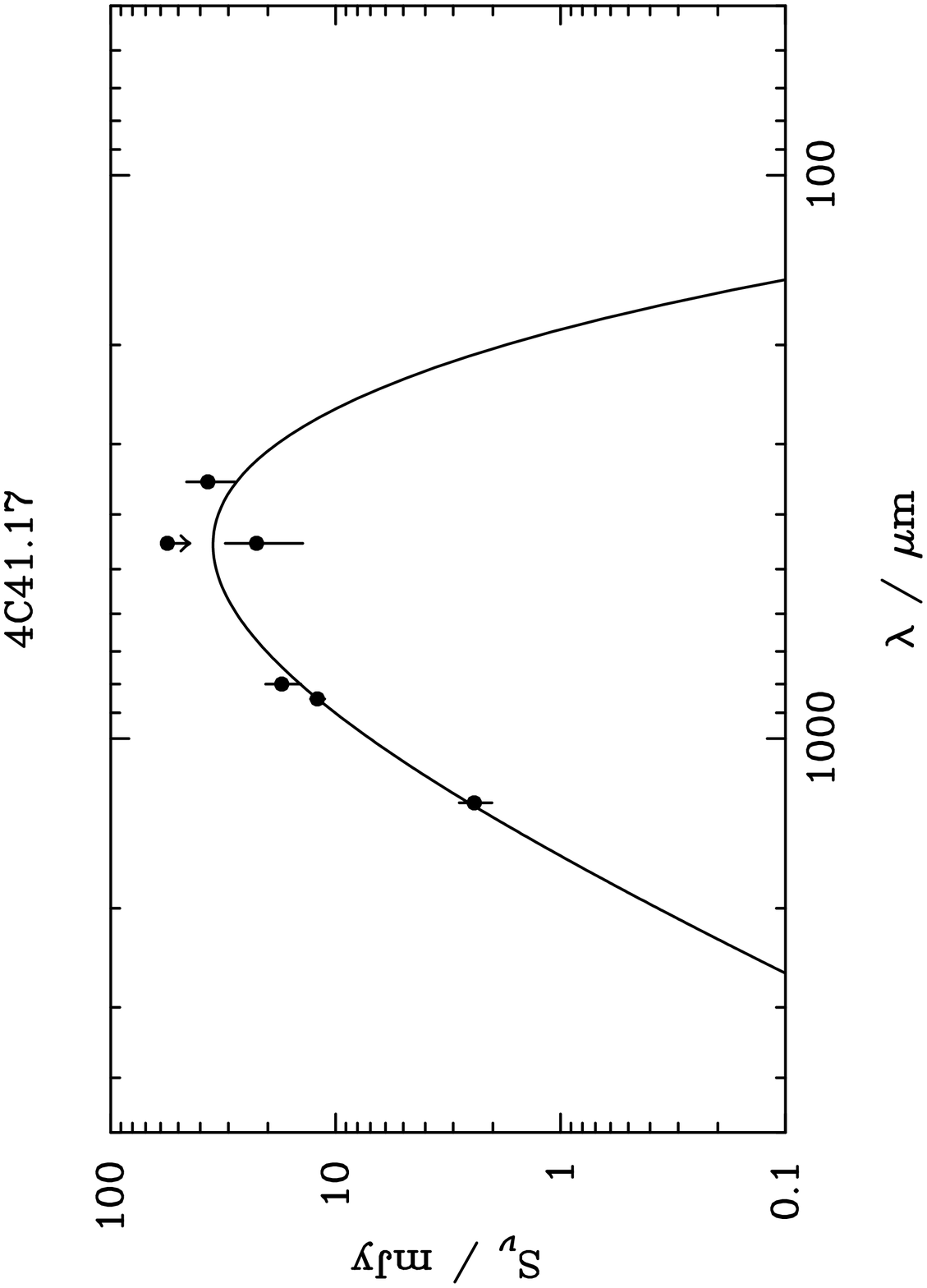}
\includegraphics{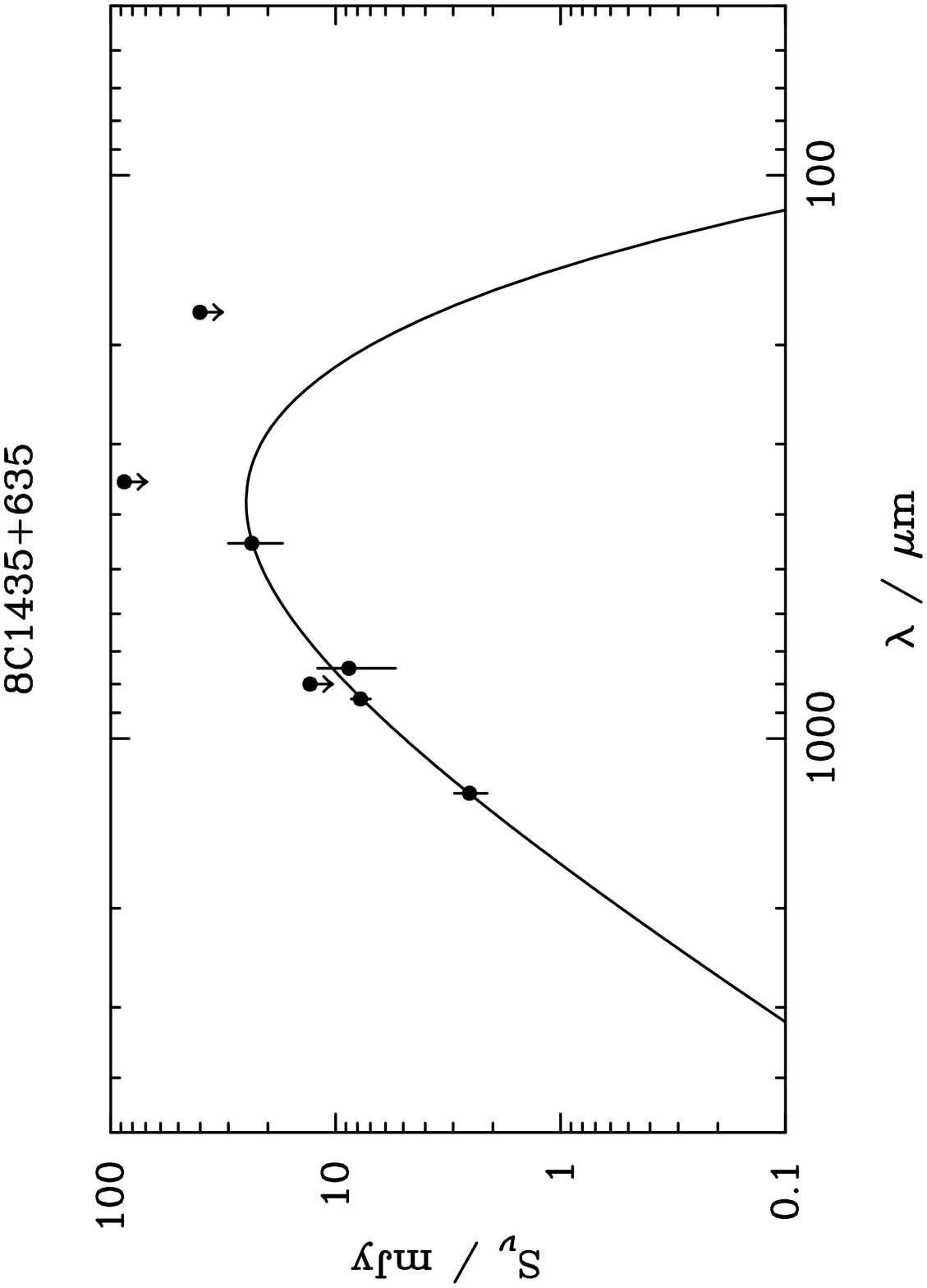}
\caption{The mm-submm spectral energy distributions of the
$z \simeq 4$ radio galaxies 4C41.17 ($z = 3.8$) and 8C1435+635 ($z = 4.25$)
as determined from a combination of observations made with the JCMT
(Dunlop et al. 1994; Ivison et al. 1998), the
IRAM 30-m telescope (Ivison 1995), and the Caltech Submm Observatory
(Benford et al. 1999). Both sources show very similar levels of sub-mm
emission $S_{850} \simeq 10$mJy, and it is clear from these SEDs that
this emission arises from dust radiating at a (rest-frame)
temperature of $T \simeq 40$K.
In both cases the inferred dust mass is $M > 10^8 {\rm M_{\odot}}$
and the inferred star-formation rate is $> 1000 {\rm M_{\odot} yr^{-1}}$.}
\end{figure}

\section{Redshifts of known sources now detected in the sub-mm}
\vspace*{-0.5cm}
\subsection{Radio galaxies}
\vspace*{-0.5cm}
The first radio galaxies to be detected at sub-mm/mm wavelengths were also 
the most distant and powerful radio galaxies known at the time - 
4C41.17 at z = 3.8 (Dunlop et al. 1994) and 8C1435$+$635 
at z = 4.2 (Ivison et al. 1995). Even before SCUBA it was clear that these
$z \simeq 4$ detections were telling us something important because attempts
to detect other well-known radio galaxies at lower redshift were much less
successful (Hughes et al. 1997). However, it was
only with the advent of SCUBA that it became possible to observe sufficiently
large samples of radio galaxies to determine if
the large dust masses found in 4C41.17 and 8C1435$+$635 (Figure 1) are
primarily a consequence of their extreme redshifts or their extreme 
radio luminosities.

My colleages and I have now completed the first such major sub-mm survey
of radio galaxies, using SCUBA in photometry mode to undertake sensitive
850-micron observations (rms noise $< 1$ mJy) of $\simeq 50$ powerful radio
galaxies spanning the redshift range $1 < z < 5$ (Archibald et al.
2000). As shown in Figure 2, this study has confirmed the
increased sub-mm detectability of radio galaxies with 
increasing redshift (15\% success rate at $z < 2.5 $, 75\% success rate at
$z > 2.5$). Moreover, we have been able to break the degeneracy between
radio luminosity and redshift, and find that the primary dependence of
sub-mm luminosity is with redshift - specifically we find that average
sub-mm luminosity increases approximately $\propto (1+z)^3$ out to 
$z \simeq 4$.
The median redshift of the sub-mm detected radio galaxies in this sample is
$z_{med} = 3.1$. It is important to stress that this is not simply a reflection
of the redshift distribution of the parent sample, as this has $z_{med} = 2$
(see Figure 2).
Interestingly, all but one of the radio galaxies with $S_{850} > 5$mJy lie at 
$z > 3$, and the median redshift for this subset is $z_{med} = 3.5$.\\
{\bf Redshift Summary:} Median $z \simeq 3.1$.\ \ 1/9 4$\sigma$
$S_{850} > 4$mJy sources at $z < 2$.\\
{\bf Bias:} Potentially unbiassed if radio galaxies are representative of 
massive galaxies in general.\\

\begin{figure}
\vspace{7.5cm}
\centering
\setlength{\unitlength}{1mm}
\includegraphics{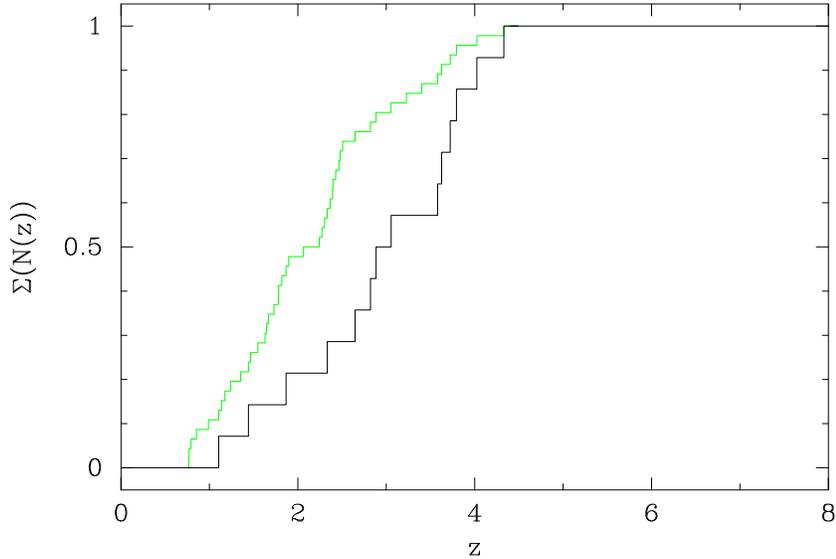}
\caption{The cumulative redshift distribution of the radio-galaxy sample
observed by Archibald et al. (2000). The green/dotted line is the entire
sample, while the black line is only for galaxies detected at 850$\mu m$. 
The median redshift of the parent sample is $z_{med} = 2$. For 
objects detected at 850-$\mu m$ the median redshift is $z_{med} = 3.1$.}

\end{figure}

\subsection{Quasars}
\vspace*{-0.5cm}
A substantial number of high-redshift ($z \simeq 4$) 
optically-selected quasars have now been detected at sub-mm 
(McMahon et al. 1999) and mm (Carilli et al. 2000) wavelengths.
However, to date such observations have been deliberately concentrated
on very high-redshift quasars, and no study has yet been completed of 
the sub-mm luminosity of QSOs of comparable optical
luminosity over a wide range in redshift, analogous to that described above
for radio galaxies. As a result it is not yet possible to quantify the
redshift dependence of dust mass in quasars. However, McMahon et al. (1999) 
have explored whether, at high redshift, there is any evidence that sub-mm
luminosity depends on optical luminosity and do not find any 
support for such a luminosity dependence.\\
{\bf Redshift Summary:} Median $z \simeq 4.4$.\ \  0/9 sources at $z < 2$.\\
{\bf Bias:} Biassed - bias potentially severe and not yet quantified.\\

\subsection{Weak radio sources}
\vspace*{-0.5cm}
Barger et al. (2000) have found that a substantial
fraction ($\simeq 1/3$) of optically-faint micro-Jy radio sources
can be detected by SCUBA at flux densities 
$S_{850} > 6$mJy. This is an important, albeit not entirely unexpected,
result as both synchrotron emission from Type II supernovae and thermal 
emission from warm dust are symptomatic of very recent
($< 10^8$yr) massive star-formation activity. Barger et al. conclude 
that $\simeq 70$\% of the bright sub-mm population can be located via 
micro-Jy radio maps, and that the entire sub-mm population lies in the
redshift band $1 < z < 3$. However, a closer examination of their data
suggests that this conclusion may be premature. In fact,
of the 4 sub-mm sources they have discovered with $S_{850} > 8$mJy, only
two are $\mu$Jy radio sources, and the listed radio-submm SED-based
redshift estimates for these 4 sources are $z = 1.5$, $z = 2.5$, $z > 3.5$
and $z > 5$.  This `survey' thus arguably
supports a mean/median redshift of $z \simeq 3$ for the bright sub-mm 
population, albeit with very large errors. In fact, even at somewhat 
lower sub-mm flux densities, examination of their figure 2 suggests that
$\mu$Jy radio-source selection can only identify $\simeq 50$\% of the SCUBA
population, especially when one factors in the inevitable bias introduced 
by imaging only sub areas which contain the majority of faint 
radio sources. Furthermore, the $\simeq 50$\% of the
SCUBA population which can be detected by this approach must, inevitably, 
be the lower-redshift half.
Wide area blank-field surveys, such as the 8-mJy survey described below,
offer the only way to properly quantify this bias.\\
{\bf Redshift Summary:} Median $z > 2.9$\ \ 1/4 sources at $z < 2$.\\
{\bf Bias:} Biased to lower-$z$ sources, but extent of bias 
not yet properly quantified.\\

\section{Redshifts of sub-mm selected sources}
\vspace*{-0.5cm}
I now consider our current information on the redshifts of sources which
have been {\it discovered} through sub-mm imaging. Such sources have been 
uncovered via three different classes of sub-mm survey, namely i) surveys 
designed to take advantage of the gravitational lensing provided by rich 
galaxy clusters, ii) surveys of the fields around known high-redshift AGN, 
and iii) unbiassed blank-field surveys, generally conducted in regions
of sky which have already been well-studied at other wavelengths.
Below I give a summary of the main results from each class of survey. I 
have deliberately focussed attention on the brightest sources, because
it is generally only for these sources that solid, unconfused sub-mm/mm 
positional information has been obtained so far.
 
\subsection{Lensing surveys}
\vspace*{-0.5cm}
The ultimate goal of spectroscopic redshifts, confirmed by CO detections,
has been achieved for three of the sources detected via SCUBA imaging of the
fields of rich clusters -- SMMJ02399$-$0136 at $z = 2.80$,
SMMJ14011+0252 at $z = 2.55$, and SMMJ02399$-$0134 at $z = 1.06$ (Smail
et al. 2000).
However, while it is obviously nice to have some concrete numbers, it is 
unclear how representative these sources are, given that their redshift 
determination was assisted by their unusually bright optical
counterparts and, in the latter two cases, by the presence of an AGN. For
the sample as a whole one has to resort to the radio-submm redshift estimator
discussed above. Based on the fact that only $\simeq 1/2$ of the sample has
proved detectable at 1.4GHz, Smail et al. (2000) argue that
the median redshift is most likely $\simeq 3$. This result 
provides direct support for the argument advanced above (in section 2.3) 
that targeted sub-mm observations of known micro-Jy radio sources are 
likely to reveal only the low-redshift half of the sub-mm galaxy population.
It also means that the median redshift of the lensing-survey sources
is in excellent accord with the
median redshift of sub-mm detected radio galaxies discussed in section 2.1.\\
{\bf Redshift Summary:} Median $z = 3$.\ \ 2/7 sources at $z < 2$.\\
{\bf Bias:} Unbiased.

\subsection{AGN companions}
\vspace*{-0.5cm}
Sub-mm surveys of the fields around known high-redshift objects are, by 
definition, inevitably biassed to some extent. However, given the 
probability that galaxy 
formation at high redshift is itself biassed towards regions of 
high density, such targeted imaging surveys are a potentially important 
complement to genuinely unbiassed blank-field surveys. Indeed the first
few deep SCUBA images of the fields around high-redshift radio galaxies 
have proved remarkably successful at uncovering bright sub-mm sources,
as illustrated in Figure 3. Deep radio imaging in the vicinity of 
such powerful radio sources is obviously challenging, but is underway at the 
time of writing. However, existing 450:850 ratios, and the faintness of 
possible optical/IR identifications already support the 
statistically-probable conclusion that these sources lie at the same 
redshift as the central
AGN. Moreover, direct confirmation of this has recently been obtained 
for one source via CO imaging (Ivison 2001). The best estimate of the 
yield of this 
work to date is 8 sources with $S_{850} > 6$mJy -- 3 at $z = 3.8$,
2 at $z = 4.4$ and 3 at $z = 3.5$.\\
{\bf Redshift Summary:} Median $z = 3.8$.\ \ 0/8 sources at $z < 2$.\\
{\bf Bias:} Biassed - imaging largely confined to objects at $z > 3$.

\begin{figure}
\vspace{7cm}
\centering
\setlength{\unitlength}{1mm}
\includegraphics{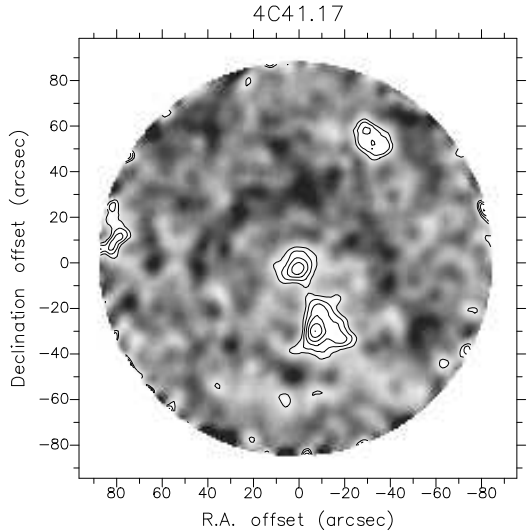}
\caption{SCUBA image of the field around the $z = 3.8$ radio galaxy
4C41.17, revealing 3 bright sub-mm sources in addition to the radio
galaxy itself (Ivison et al. 2000).}
\end{figure}

\subsection{Blank-field surveys}
\vspace*{-0.5cm}
\subsubsection{The Hubble Deep Field}
\vspace*{-0.2cm}
At the time of writing none of the 
sub-mm sources in the Hubble Deep Field reported by Hughes et al. (1998) has 
a reliable redshift. Given the wealth of supporting data this serves to
demonstrate how hard it will be to obtain complete redshift information on 
sub-mm selected samples. Interesting constraints can, however, be placed 
on the redshift of the brightest source HDF850.1 ($S_{850} = 7$mJy), following 
its detection with the IRAM PdB interferometer at 1.3mm (Downes et al.
1999). While the accurate position provided by the interferometric detection
has not in fact yielded an unambiguous optical identification, it has enabled
a robust upper limit to be placed on the 1.4GHz:850$\mu m$ flux-density ratio
of $> 200$. This, along with the 450:850 and 850:1300 ratios all point
to a redshift in the range $3 < z < 5$ (see Figures 4 and 5), in which 
case it is certainly possible that its optical/IR counterpart is too faint to
be detected by existing optical/IR imaging. Radio-submm SED constraints
suggest $z \simeq 2$ for the second-brightest source in the HDF.\\
{\bf Redshift Summary:} Median $z = 3$.\ \ 0/2 sources at $z < 2$.\\
{\bf Bias:} Unbiassed, but sample very small.\\

\begin{figure}
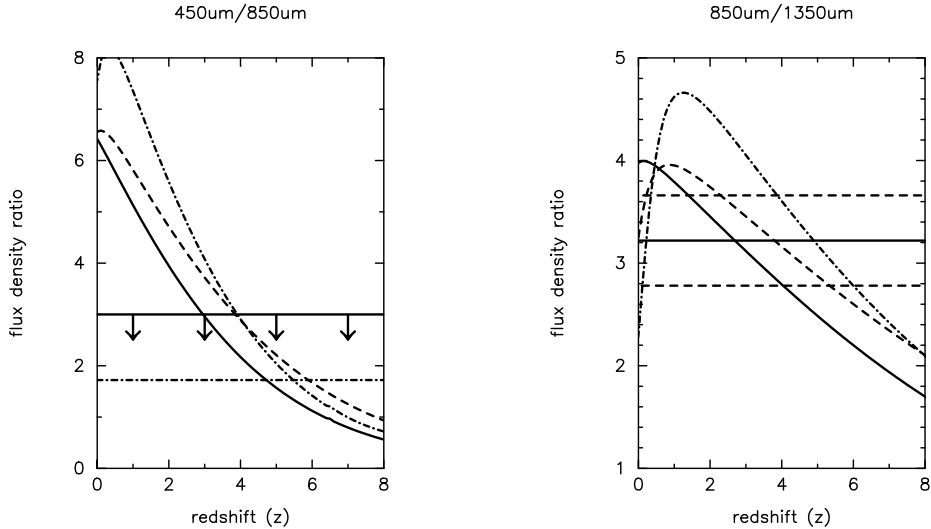

\vspace{7cm}
\leavevmode
\includegraphics{firsedfig4a.eps}
\includegraphics{firsedfig4b.eps}
\caption{Observed 450/850$\mu$m and 850/1350$\mu$m flux-density ratios
for HDF850.1 compared with what might be expected  as a function of
redshift. The curves show the extreme range of colours for SEDs that
represent dust enshrouded starburst galaxies and AGN - Arp220 (solid),
M82 (dashed) and Mkn231 (dot-dashed). The solid horizontal line in the
left-hand panel shows the existing  upper limit to the 450/850$\mu$m
flux ratio for HDF850.1. The horizontal lines in the RH panel show
the measured 850/1350$\mu$m flux ratio for HDF850.1 (solid line), and the
$\pm 1\sigma$ errors on this ratio (dot-dashed lines).}
        
\end{figure}

\subsubsection{The Canada-UK Deep Sub-mm Survey}
\vspace*{-0.2cm}
The recent completion of the 14-hr field of the Canada-UK SCUBA survey 
has yielded 7 sources with signal:noise $> 4 \sigma$ and
$S_{850} > 4$ mJy (Eales et al. 2000).
Of particular interest is the brightest source 
(CUDSS 14.1; $S_{850} =  9$ mJy) which, like HDF850.1 has now been detected 
with the the IRAM PdB interferometer at 1.3mm (Gear et al. 2000). Again 
radio-mm-submm SED constraints
favour a high redshift in the range $2 < z < 4.5$; Eales et al.
choose an estimated redshift of $z \simeq 2.2$. Most of the other 
4-$\sigma$ sources in this sample have not been detected at 1.4GHz, but 
their relatively
weak sub-mm fluxes means the resulting redshift constraint is generally
simply $z > 2$.\\
{\bf Redshift Summary:} Median $z > 2.1$.\ \ 1/7 sources at $z < 2$.\\
{\bf Bias:} Unbiassed, but only 3 sources brighter than 5 mJy, so
radio-submm redshift constraints are weak.\\

\subsubsection{The 8-mJy survey}
\vspace*{-0.2cm}
The ongoing SCUBA survey of the ELAISN2 and Lockman-E fields has now 
covered 150 sq. arcmin to even depth, and has currently yielded 12 sources 
with signal:noise $> 4 \sigma$ at flux densities $S_{850} > 8$mJy
(Scott et al. 2001). Very deep radio data will not
be available for these fields until early in 2001, and so at present it is
not possible to place even crude constraints on the redshifts of most 
of these sources. However, the brightest source in the Lockman field has 
been detected by the IRAM PdB interferometer, and in this case identified 
with a faint red complex object via very-deep $K$-band imaging. As shown 
in Figure 5, both SED
constraints and the properties of this infrared identification
favour a redshift in the range $2 < z < 4.5$ (Lutz et al. 2001; Dunlop 2001).\\
{\bf Redshift Summary:} Median $z \simeq 3$.\ \ 0/2 sources at $z < 2$.\\
{\bf Bias:} Unbiassed, but currently lacking in redshift information.
Addition of deep radio data in early 2001 will transform this situation.\\

\begin{figure}
\vspace{6.2cm}
\centering
\setlength{\unitlength}{1mm}
\includegraphics{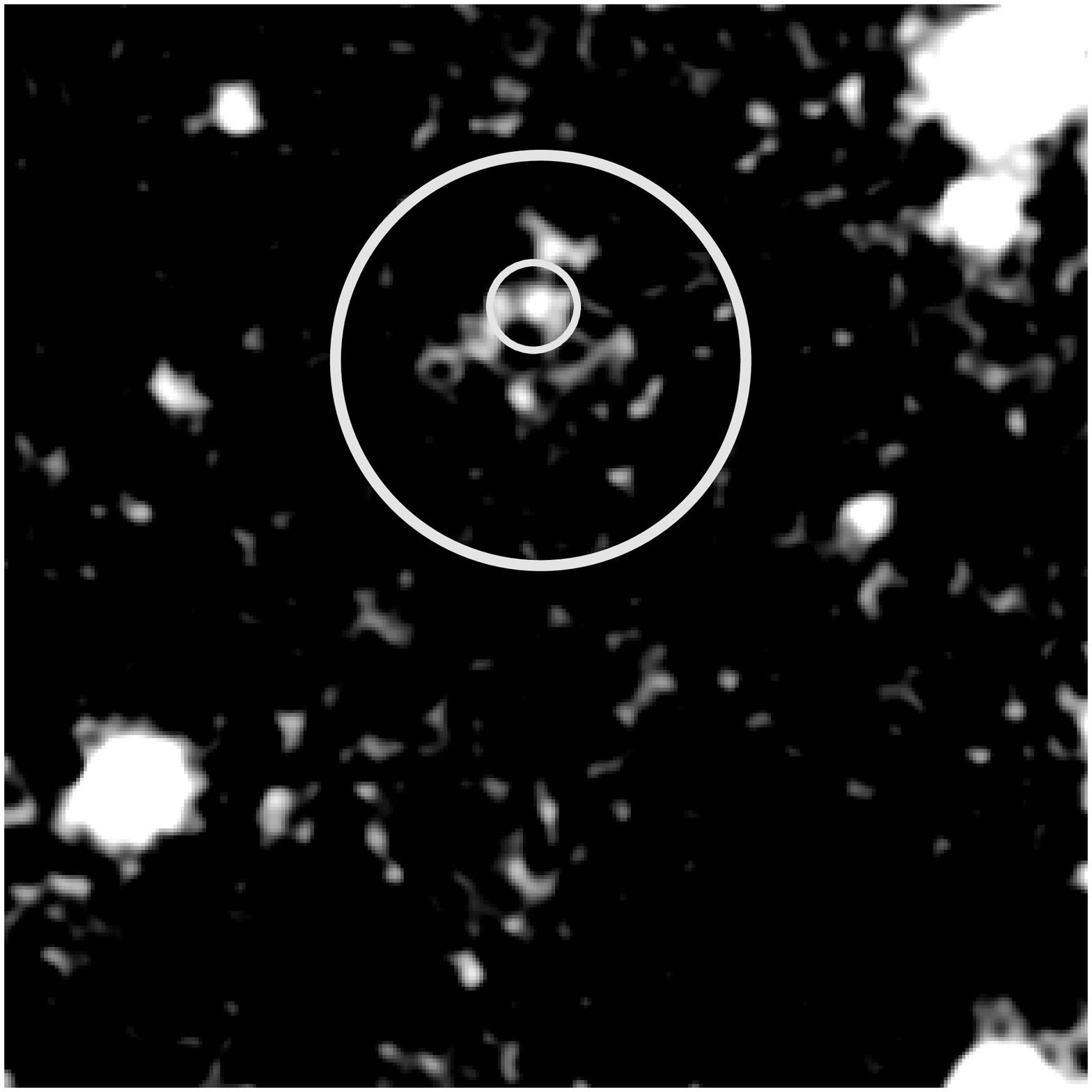}
\includegraphics{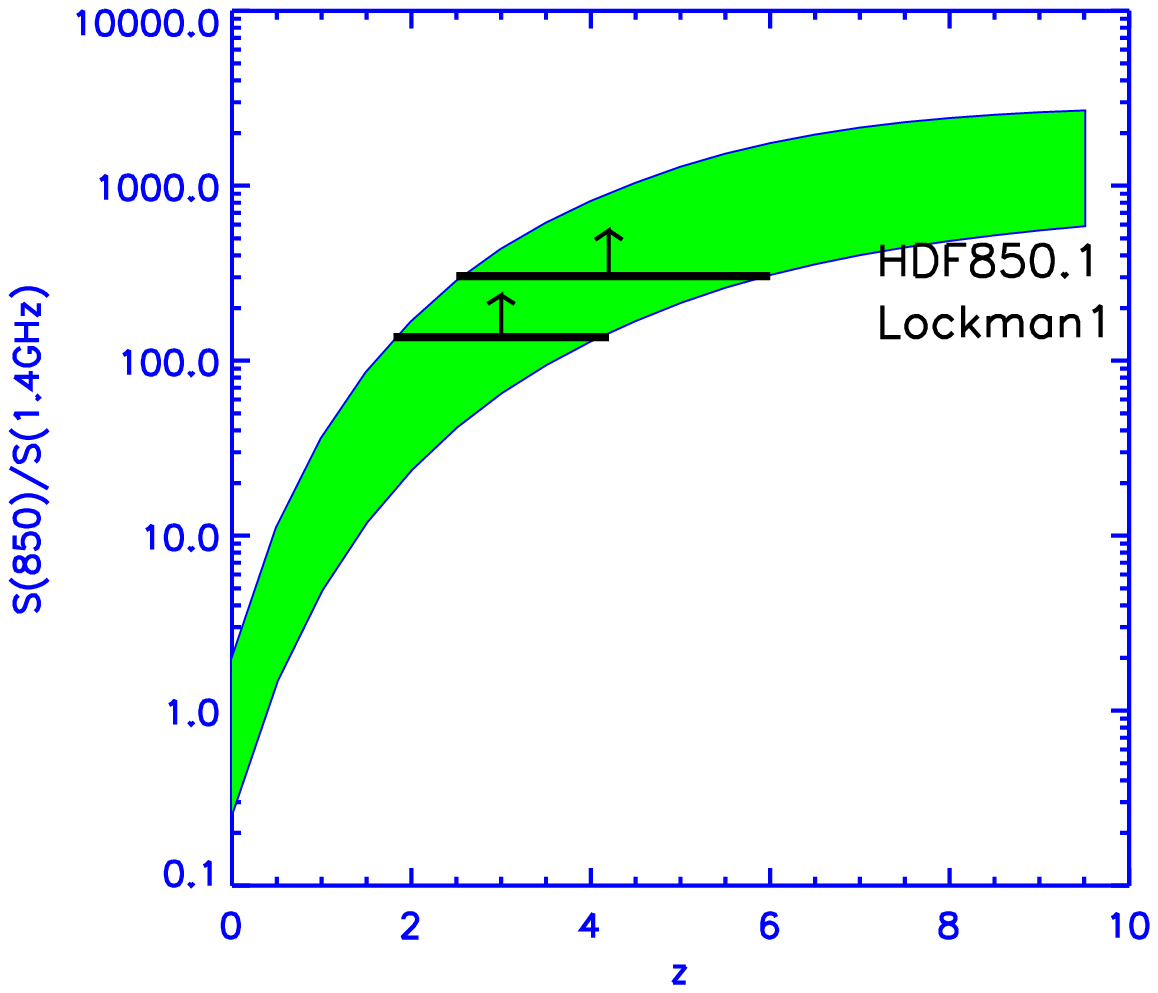}
\caption{Current information on the brightest SCUBA source in the 
Lockman field of the SCUBA 8-mJy survey. The left-hand panel shows a 30 x 30
arcsec region of a deep (6-hr on UKIRT) K-band image of the
field in the vicinity of the $S_{850\mu m} = 11$ mJy source Lockman850.1.
The position of the SCUBA source is marked by the large (10-arcsec
diameter) circle, while the position of the 1.3mm source detected in
follow-up observations with the IRAM PdB interferometer is marked by the
small (2-arcsec diameter) circle.
With this positional accuracy the SCUBA source
can be confidently identified with the brightest of a group of
compact peaks with $K \simeq 21$, and $R-K > 5$ (Lutz et al. 2001). 
The right-hand panel 
shows the current radio-mm SED redshift constraints on the redshift of
this source, along with the analogous constraints on HDF850.1
for which deeper radio data are currently available.}
\end{figure}

\label{}

\section{Conclusion}
\vspace*{-0.5cm}
The results discussed above are summarized in Table 1.
Despite the biases inherent in several of these surveys, it is 
interesting that of the $\simeq 50$ significant extragalactic
sources with $S_{850} > 4$mJy uncovered by SCUBA during its first 2-3 years of 
operation, only $\simeq 10$\% appear to lie at $z < 2$. Certainly, 
all of these studies can be reconciled with a median redshift of at least 
$z_{med} = 3$ for the luminous sub-mm galaxy population.

Such a high median redshift is arguably not unexpected given current 
theories designed to explain the correlation between black-hole mass and 
spheroid mass found at low redshift. In such scenarios, peak AGN emission is 
expected to correspond to, or even to cause termination of major 
star-formation 
activity in the host spheroid. In contrast, maximum dust emission is expected 
to occur approximately half-way through the star-formation process
(Eales \& Edmunds 1996; Archibald et al. 2001). Given that
the optical emission from powerful quasars peaks at $z \simeq 2.5$, 
dust-emission 
from massive ellipticals might be expected to peak 0.5-1 Gyr earlier, at 
$z \simeq 3$. Confirmation or refutation of this appealingly-simple picture 
requires substantially-improved redshift 
information on bright samples of SCUBA sources,
such as the 8mJy survey.

\begin{table}
\begin{centering}
\begin{tabular}{|lr|c|r|}
\hline
Survey & No. & No. at $z < 2$ & $z_{median}$ \\
\hline
Radio Galaxies  & 9 & 1 & 3.1 \\
Quasars         & 9 & 0 & 4.4 \\
$\mu$-Jy Radio  & 4 & 1 & $>$2.9\\
Lensing Surveys & 7 & 2 & 3\\
AGN fields      & 8 & 0 & 3.8\\
HDF             & 2 & 0 & 3\\
CUDDS           & 7 & 1 & $>$2.1\\
8mJy            & 2 & 0 & 3\\
\hline
Total           & 48 & 5 & \\
\hline
\end{tabular}
\caption{Our current knowledge of the redshifts 
of sources with $S_{850} > 4$mJy and S/N $>$ 4.}
\end{centering}
\end{table}

\section*{Acknowledgments}
\vspace*{-0.5cm}
I gratefully acknowledge the contributions of my collaborators on several
of the projects covered in this review, especially Elese Archibald, 
Dave Hughes, Rob Ivison, Steve Rawlings, Omar Almaini, Dieter
Lutz, John Peacock, Suzie Scott, Steve Serjeant and the 
other members of the UK SCUBA Consortium.

\section*{References}
\vspace*{-0.5cm}
Archibald, E.N., et al., 2000, MNRAS, in press (astro-ph/0002083).\\
Archibald, E.N., Dunlop, J.S., et al., 2001, ApJ, in preparation.\\
Barger, A.J., Cowie, L.L., Richards, E.A., 2000, AJ, 119, 2092.\\
Benford, D.J., et al., 1999, ApJ, 518, L65.\\
Blain, A.W., 2001, In: `{\it Deep Sub-millimetre Surveys}', eds. Lowenthal, J.
\&\\
\hspace*{0.5cm} Hughes, D., World Scientific, in press (astro-ph/0009012).\\
Blain, A.W., Smnail, I., Ivison, R.J., Kneib, J.-P., 1999, MNRAS, 302, 632.\\
Carilli, C.L., Yun, M.S., 1999, ApJ, 513, L13.\\
Carilli, C.L., Yun, M.S., 2000, ApJ, 530, 618.\\
Carilli, C.L., et al., 2000, ApJ, in press (astro-ph/0002386).\\
Downes, D., et al, 1999, A\&A, 347, 809.\\
Dunlop, J.S.,  et al., 1994, Nature, 370, 347.\\
Dunlop, J.S., 2001, In: `{\it Deep Sub-millimetre Surveys}', eds. Lowenthal, J.
\&\\
\hspace*{0.5cm} Hughes, D., World Scientific, in press (astro-ph/0011077).\\
Dunne, L., Clements, D.L., Eales, S.A., 2000, MNRAS, 315, 115.\\
Eales, S.A., Edmunds, M.G., 1996, MNRAS, 280, 1167.\\
Eales, S.A., et al., 2000, AJ, in press (astro-ph/0009154).\\
Gear, W.K., et al., 2000, MNRAS, in press (astro-ph/0007054).\\
Hughes, D.H., Dunlop, J.S., Rawlings, S., 1997, MNRAS, 289, 766.\\
Hughes, D.H.,  et al., 1998, Nature, 394, 241.\\
Ivison, R.J., 1995, MNRAS, 275, L33.\\
Ivison, R.J., 2001, In: `{\it Deep Sub-millimetre Surveys}', eds. 
Lowenthal, J. \&\\
\hspace*{0.5cm} Hughes, D., World Scientific, in press.\\
Ivison, R.J.,  et al., 2000, ApJ, in press (astro-ph/0005234).\\
Lutz, D., et al., 2001, A\&A, in preparation.\\
McMahon, R.G., et al., 1999, MNRAS, 309, L1.\\
Scott, S., et al., 2001, MNRAS, in preparation.\\
Smail, I., et al., 2000, ApJ, 528, 612.\\

\end{document}